\documentclass[sigconf]{acmart}

\usepackage[tight,footnotesize]{subfigure}

\newtheorem{example}{Example}[section]

\newtheorem{situation}{Situation}[section]

\AtBeginDocument{%
  \providecommand\BibTeX{{%
    \normalfont B\kern-0.5em{\scshape i\kern-0.25em b}\kern-0.8em\TeX}}}

\copyrightyear{2021} 
\acmYear{2021} 
\setcopyright{acmcopyright}\acmConference[UbiComp-ISWC '21 Adjunct]{Adjunct Proceedings of the 2021 ACM International Joint Conference on Pervasive and Ubiquitous Computing and Proceedings of the 2021 ACM International Symposium on Wearable Computers}{September 21--26, 2021}{Virtual, USA}
\acmBooktitle{Adjunct Proceedings of the 2021 ACM International Joint Conference on Pervasive and Ubiquitous Computing and Proceedings of the 2021 ACM International Symposium on Wearable Computers (UbiComp-ISWC '21 Adjunct), September 21--26, 2021, Virtual, USA}
\acmPrice{15.00}
\acmDOI{10.1145/3460418.3480412}
\acmISBN{978-1-4503-8461-2/21/09}

\begin{document}

\title{Data-driven Clustering in Ad-hoc Networks based on Community Detection}

\author{Shufan Huang }

\affiliation{%
  \institution{Shanghai Jiao Tong University}
  \country{China}
}
\email{hsf19991102@sjtu.edu.cn}

\author{Yongpeng Wu}
\affiliation{%
  \institution{Shanghai Jiao Tong University}
  \country{China}
  }
\email{yongpeng.wu@sjtu.edu.cn}
\authornote{Corresponding author.}

\author{Siyuan Gao }
\affiliation{
  \institution{Jiangxi Institute of Metrology \& Testing}
  \country{China}
  }
\email{37531309@qq.com}



\begin{abstract}
High demands for industrial networks lead to increasingly large sensor networks. However, the complexity of networks and demands for accurate data require better stability and communication quality. Conventional clustering methods for ad-hoc networks are based on topology and connectivity, leading to unstable clustering results and low communication quality. In this paper, we focus on two situations: time-evolving networks, and multi-channel ad-hoc networks. We model ad-hoc networks as graphs and introduce community detection methods to both situations. Particularly, in time-evolving networks, our method utilizes the results of community detection to ensure stability. By using similarity or human-in-the-loop measures, we construct a new weighted graph for final clustering. In multi-channel networks, we perform allocations from the results of multiplex community detection. Experiments on real-world datasets show that our method outperforms baselines in both stability and quality.
\end{abstract}




\maketitle

\section{Introduction}
\label{sec:intro}

The development in the industry led to higher demand on management of energy consumption data. 
The number and variety of data require a huge number of sensors, and the connection between the sensors would be a problem to be solved. As the number of sensors increases, the network becomes more complex.
In ad-hoc networks, clustering algorithms 
separate the nodes in ad-hoc networks into multiple groups and determine the connection between nodes. They also enable the ad-hoc networks to adjust according to changes of situation.

However, conventional clustering methods in ad-hoc networks are mainly based on topology and connectivity, such as k-hop and dominant set-based clustering algorithms \cite{cluter_survey2}. Although clustering algorithms are efficient, the clustering quality cannot be guaranteed. In sensor networks, the quality of data affect the subsequent analysis.
Thus, new methods should facilitate further analysis. Furthermore, the conventional clustering is only based on the current status of the ad-hoc network. Results could be unstable due to the lack of previous data. Also, in ad-hoc networks with multiple channels, conventional methods cannot map the sensors to each channel and maximize the quality of communication at the same time.

In this paper, we propose two methods to solve the clustering of ad-hoc networks under two different situations by modelling the ad-hoc network as graphs. In the first situation, behaviors of ad-hoc networks over a period of time is recorded and time-evolving graphs are constructed. Our objective is to make use of the historical data in the ad-hoc network to generate a stable and reliable partition of the graph. The second situation can be depicted as a multiplex graph, where layers in the graph denote different channels in the ad-hoc network. By assigning labels for each node in the graph, the nodes can be clustered and allocated to their optimal channel.
Inspired by the social property in graphs generated from ad-hoc networks, we introduce two methods based on community detection. The first method is for the time-evolving graphs, where time series of graphs are treated independently at each timestamp. Then, we treat the community relationships between nodes in a period of time as the basis for the reconstruction of the graph. We offer two options: the first is to calculate an index based on the Jaccard Similarity as the weight of new edges; the other is to train a classifier with a small scale of labeled data to output the new weight of the edges. 
The second method is for ad-hoc networks with multiple channels, a mapping of nodes can be established by labeling sensors instead of nodes in the multiplex network. With the mapping, we further find the optimal way to allocate the sensors to channels. In summary, this paper made the following contributions:
\begin{itemize}
    \item We introduce the community structure from social network to model the structure of the ad-hoc network.
    \item We propose two methods based on data-driven community detection suitable for time-evolving ad-hoc networks and multi-channel ad-hoc networks.
    \item The experiment shows that the proposed methods outperform conventional clustering algorithms in both stability and communication quality.
\end{itemize}

The rest of the paper is organized as follows. In Section \ref{sec:related}, we review the related works regarding conventional clustering methods for ad-hoc networks and community detection methods. Section \ref{sec:model} introduces our key motivation and models for two situations. In Section \ref{sec:method}, we propose our novel methods based on community detection. In Section \ref{sec:exp}, we introduce the experimental setup, analyze the results of different methods and discuss the interesting findings. Finally, we conclude remarks and future works in section \ref{sec:conclusion}.

\section{Related Works}
\label{sec:related}

\subsection{Clustering Methods for Ad-hoc Networks}
\label{sec:related:cluster}
The existing clustering algorithms for ad-hoc networks are mostly unsupervised algorithms based on graph data, including algorithms based on topology, multi-hop clustering, and weighted clustering.
Some clustering algorithms are defined based on topology, such as the concept of dominating set \cite{cluter_survey2}. Commonly used dominating sets includes Connected Dominating Set (CDS) \cite{CDS} and Weakly Connected Dominating Set (WCDS) \cite{WCDS}. These algorithms define clusters as dominating sets or dominating set with weaker definitions. Such algorithms are easy to implement but the efficiency is restricted since the discovery of dominating set is an NP-hard problem.
In multi-hop clustering, the clusters are generated with the information of the neighborhood. The cluster heads are selected based on their own IDs and the IDs of their k-hop neighbors. The nodes communicate with each other by broadcasting their IDs and decisions of selection, an example of such clustering methods is the  k-CONID \cite{kconid}, it is a combination of multi-hop clustering and identifier neighbor based clustering \cite{clustering_survey}.
Weighted clustering is the basis of many commonly used clustering algorithms in ad-hoc networks, such as the Weighted Clustering Algorithm (WCA) \cite{WCA}. The greatest advantage of weighted clustering algorithms is their flexibility. A weighted combination of attributes such as node degree, node mobility, transmission range, distance to neighbors is used to score the nodes. The cluster head is elected by broadcasting the scores among neighbors.

\subsection{Community Detection}
\label{sec:related:community}

Community detection is a useful tool in the analysis of large-scale networks. All the community detection methods are based on the community structure in the network. The community structure refers to a group of nodes in the graph whose density is large internally, while the external connections are sparse. Various community detection algorithms have been proposed with different objective functions, such as modularity-based algorithms (Louvain) \cite{Louvain}, information entropy-based algorithm map equation (InfoMap) \cite{infomap}, connectivity based algorithms (GN) \cite{GN}, label propagation algorithms (LPA) \cite{LPA},  non-negative matrix factorization \cite{NMF}, and methods based on compressed sensing \cite{CS1}\cite{CS2}. The community structure allows the division of large-scale networks and makes the subsequent analysis easier.
For community detection in multilayer graphs, the methods can be categorized into three types. \cite{survey1} The first is flattening methods, which collapses the information in a multilayer graph into a single layer. These methods are commonly used in the multiplex networks, where inter-layer edges only appears between same nodes. The second is aggregation methods, which discovers communities in each layer, and then aggregates them by a certain principle. Direct methods are community algorithms that optimizes some quality-assessment criteria without flattening, such as Locally Adaptive Random Transitions(LART) \cite{LART}, Label Propagation Algorithm for the multilayer network (MNLPA) \cite{MNLPA}, MultiTensor \cite{MultiTensor} and InfoMap \cite{infomap_multi}.


\section{Motivation and Modeling}
\label{sec:model}
In this section, we explain the motivation and model to solve the problem of clustering in ad-hoc networks.

\subsection{Motivation}
\label{sec:model:motiv}

Sensor network plays an increasingly important role in monitoring operation status and data analytics. The most significant feature of ad-hoc networks is its flexibility and adaptation to changes in the network, such as new nodes and failure in nodes. Nonetheless, the failure and insertion of sensors are not-so-common in some industrial applications. Therefore, the demand for ad-hoc networks changes from robustness to stability and the quality of communication. Conventional clustering algorithms in ad-hoc networks focus more on the connectivity and topology of the network, and new algorithms are needed to ensure the quality of communication.
Researches have been conducted on the \emph{social properties} of ad-hoc networks \cite{sociality, socialty2, socialty3}. In sensor networks, the phenomenon of social relationships is easy to identify. For example, the sensors monitoring the same object have a natural tendency to communicate with each other, which reflects the social property in the network. Therefore, as a useful tool in analysis in social networks, community detection is also applicable on ad-hoc networks.

\subsection{Model}
\label{sec:model:model}

In this paper, we set two different situations. 
\begin{situation}
\label{sit:1}
The quality of communication between sensors is given for a period of time, a method is needed to partition the sensors into smaller clusters.
\end{situation}
\begin{situation}
\label{sit:2}
The quality of communication between sensors of different channels is given, the method is required to give an allocation to maximize the quality of communication on each channel, and minimize the number of isolated sensors.
\end{situation}

We model the ad-hoc network by undirected graphs $G(V, E, W)$, where nodes $v \in V$ denote sensors, edges $e \in E$ denote communications between the two sensors, and edge weights $w \in W$ denote communication quality between the two sensors.


In Situation \ref{sit:1}, an undirected graph is created for every timestamp to represent the time-evolving network. In Situation \ref{sit:2}, each channel is viewed as a layer of the graph. Since there are multiple channels in the second situation, the network becomes a multilayer graph. Besides, since there is no interlayer connection, the graph can be further simplified into a multiplex graph, where the interlayer connection exists if and only if the two nodes have the same index in different layers.

In these two situations, we introduce community detection methods to split the original network into communities in which the nodes are densely connected to each other. 
Given an undirected weighted graph $G(V, E, W)$, community detection algorithms outputs a community assignment $C$, grouping $V$ to corresponding labels signaling the community.
An evaluation method for community assignment is needed to obtain the best result. The evaluation in Situation \ref{sit:1} is defined as an object function to compute the average modularity\cite{modularity} $\bar{Q}$ in the time-evolving graphs, which is a modified version of modularity  in multilayer networks.
\begin{equation}
    \bar{Q}=\frac{1}{T}\sum^{T}_{t=1}\sum^{c_j}_{i=1}(e_{ii}(t)-a_{i}(t)^2)
\end{equation}
where $e_{ii}$ is the fraction of the edges from one node in group $i$ to another node in the same group $i$ in the graph at timestamp $t$, $c_j$ is the number of communities at timestamp $j$. $a_{ih}$ is the expected fraction of edges from one node in group $i$ to another one in the same group in a randomized network at timestamp $t$. $T$ is the length of time window selected from the time-evolving graph. Larger modularity indicates stronger community structure. 

Besides modularity, we include Bit Error Rate (BER) to evaluate the quality of communication. With higher density in the community, the average BER calculated with the average shortest distance in each connective component in the graph is also reduced due to the increment in density when modularity is optimized.


In Situation \ref{sit:2}, the evaluation objective $L$ is composed of connectivity and average BER.  Connectivity is defined as the number of connected components in a graph where the inter-community edges are removed. Connectivity is introduced to ensure that every node can be reached by other nodes in the same community, and no extra connected components are separated from the network due to the allocation of channels. 
\begin{equation}
    L(C)=\bar{Q}+\lambda (N_{cc}-N_{community})+\gamma BER
\end{equation}
where $C$ is the final result of community detection, $N_{cc}$ is the number of connected components in a graph where the inter-community edges are removed, $N_{communities}$ is the number of communities in the final, and BER is the average Bit Error Rate between nodes. $\lambda$ and $\gamma$ are weights to balance the importance of the three factors.

\section{Methodology}
\label{sec:method}


In this section, we introduce two methods that can be adopted to serve as a clustering algorithm and give a satisfying result on real-life datasets. The first method is a combination of community detection results on multiple timestamps. It facilitates the overall analysis of sensor networks over a period of time. The second method is to adapt multi-layer community detection algorithms to the clustering of networks with multiple channels.


\subsection{Method for Situation \ref{sit:1}}
\label{sec:method:1}

The first method is to solve the problem of network partition over a period of time. The main idea is to use the statistic result as the parameter in the network structure, and reach an overall result by applying the community detection algorithms. It can be divided into four steps.

\subsubsection{Graph Construction}
\label{sec:method:1:graph}
Firstly, this method takes the communication data between sensors as input. The input data is split by timestamp. For each timestamp, the communication from one sensor to another is unique. The sensors can be viewed as nodes in the networks, and the communication is treated as undirected edges between sender and receiver. 
The network becomes an undirected graph with average communication data as the weight of corresponding edges.

\subsubsection{Community Detection}
\label{sec:method:1:community}
Secondly, community detection algorithms are employed to get the community structure at the specific timestamp. The community detection algorithm can be any algorithm that can be employed on the undirected weighted network. In this paper, we leverage Louvain \cite{Louvain}, GN \cite{GN}, and InfoMap \cite{infomap} to evaluate their performance.

\subsubsection{Graph Reconstruction}
\label{sec:method:1:recon}
Thirdly, the results generated in the second step are gathered to construct a new graph containing the information from all timestamps.
Here, two options are provided, the first is to traverse all timestamps to calculate the number of timestamps where two nodes are in the same community. A variation of Jaccard Similarity is used to represent how close the relationship is between two nodes.
\\
\begin{equation}
    J=\frac{C_t(A)\cap C_t(B)}{C_t(A)\cup C_t(B)}
\end{equation}
where $C_t$ is the set of community mapping with timestamp, and A,B are two different nodes in the graph.\\
\begin{figure*}[h]
\centering 
\includegraphics[width=1\textwidth]{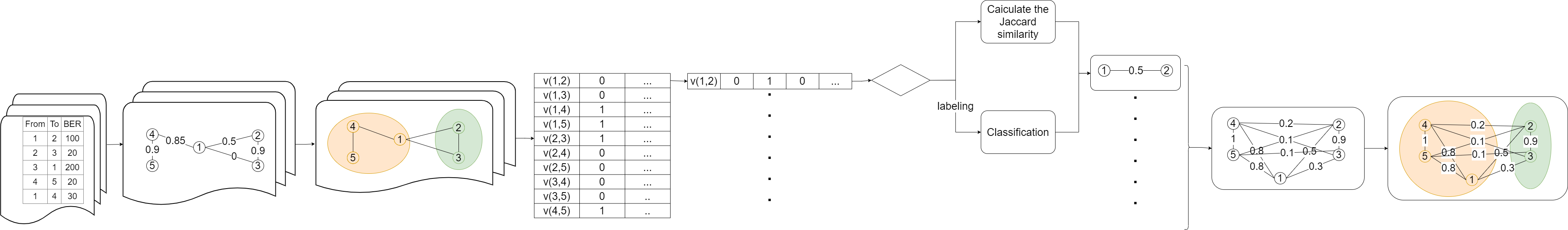}
\caption{Method Overview} 
\label{Fig.main2} 
\vspace{-1.0em}
\end{figure*}
The second option is based on a simple extreme learning machine that serves as a classifier. A binary vector over a time window is recorded for classification. Zero means that the two nodes are in the same community at the corresponding timestamp, while one stands for the opposite. If there is no prior knowledge about the vector, some candidates are proposed and people can label whether the two nodes are in the same community at the next timestamp. In the following processing, further manual labeling is not needed, since the result of community detection in the next timestamp will repeat this process. 
In the next step, a new graph is created using the ratio or probability above. All nodes that appeared in the period of time are included. The edges are non-directional edges with the ratio of its node pair as weight. The new network usually contains more edges than the network of any timestamp. The ratio serves as the similarity between nodes, and the community detection algorithm can be employed again to get the final result.

\subsubsection{Node Partition}
\label{sec:method:1:eva}

Since LPA \cite{LPA} only requires communication in the neighborhood, which facilitates its implementation on distributed systems, including the ad-hoc network, it is applied on the reconstructed graph in Step \ref{sec:method:1:recon}.
The final community splits the final network into smaller parts, thus the community detection algorithm can be viewed as the clustering algorithm.

\subsection{Method for Situation \ref{sit:2}}
\label{sec:method:2}
The second method is to solve the problem of multi-channel partition. For sensor networks with multiple channels, each sensor can communicate on different channels. Each channel can be regarded as a layer in the network, the communications between nodes are denoted as edges on the correspondent layer. The sensors are represented by multiple nodes on every layer.

\subsubsection{Community Detection}
\label{sec:method:2:com}

We reuse graph construction step in Section \ref{sec:method:1:graph} but alter the community detection step to fit the situation of multi-layer.
Since there are no interlayer connections between different nodes, the network can be viewed as multiplex graphs, where interlayer connections can only be between nodes of the same sensor. \cite{kinsley2020multilayer} Conventional partition methods based on topology may find it hard to fit such situations, but with multi-layer community detection algorithms, the problem can be solved. Community detection algorithms can generate a map from the node index to its community. With this map, every node belongs to a community, and the communities can serve as partitions because they tend to cluster nodes with the greatest similarity in the multiplex graph. Thus, we leverage multi-layer community detection algorithms including LART \cite{LART}, MNLPA \cite{MNLPA}, MultiTensor \cite{MultiTensor}, and InfoMap \cite{infomap_multi} to output the community partition.





\subsubsection{Channel Distribution}
\label{sec:method:2:dis}

With multiplex communities, the next step is to select the channels suitable for each community. A greedy algorithm is used to solve this problem. The communities are sorted by the number of nodes, and the larger community has the priority to choose the most suitable channel for them based on two requirements. The first one is that if the community is set on the channel, the increment of connected components is minimized. The second one is that the communication quality within the community is maximized. For the first $n$ communities, where $n$ is the number of channels, they should be in different layers. While for the latter communities, in case two requirements collide, the first one is the priority.

\section{Experiment and evaluation}
\label{sec:exp}

In this section, the result of experiment based on real-life datasets are presented. We compare different algorithms in Methods for both Situations.
For Situation \ref{sit:1}, we compare the performance of the method with different community detection algorithms. In Situation \ref{sit:2}, we compare the performance of several multi-layer community detection algorithms.

\subsection{Dataset and Settings}
\label{sec:exp:dataset}

Table \ref{tab:freq} shows the dataset used for Situation \ref{sit:1}. It includes the data collected by a sensor network in three periods of time.
The dataset used for Situation \ref{sit:2} is from communications between sensors and their quality in three different channels during the same time period, with 62 timestamps.


\begin{figure*}[h]
\vspace{-1.0em}
\centering
\subfigure[Variation]{
\label{Fig.phase1.1}
\includegraphics[width=0.3\textwidth]{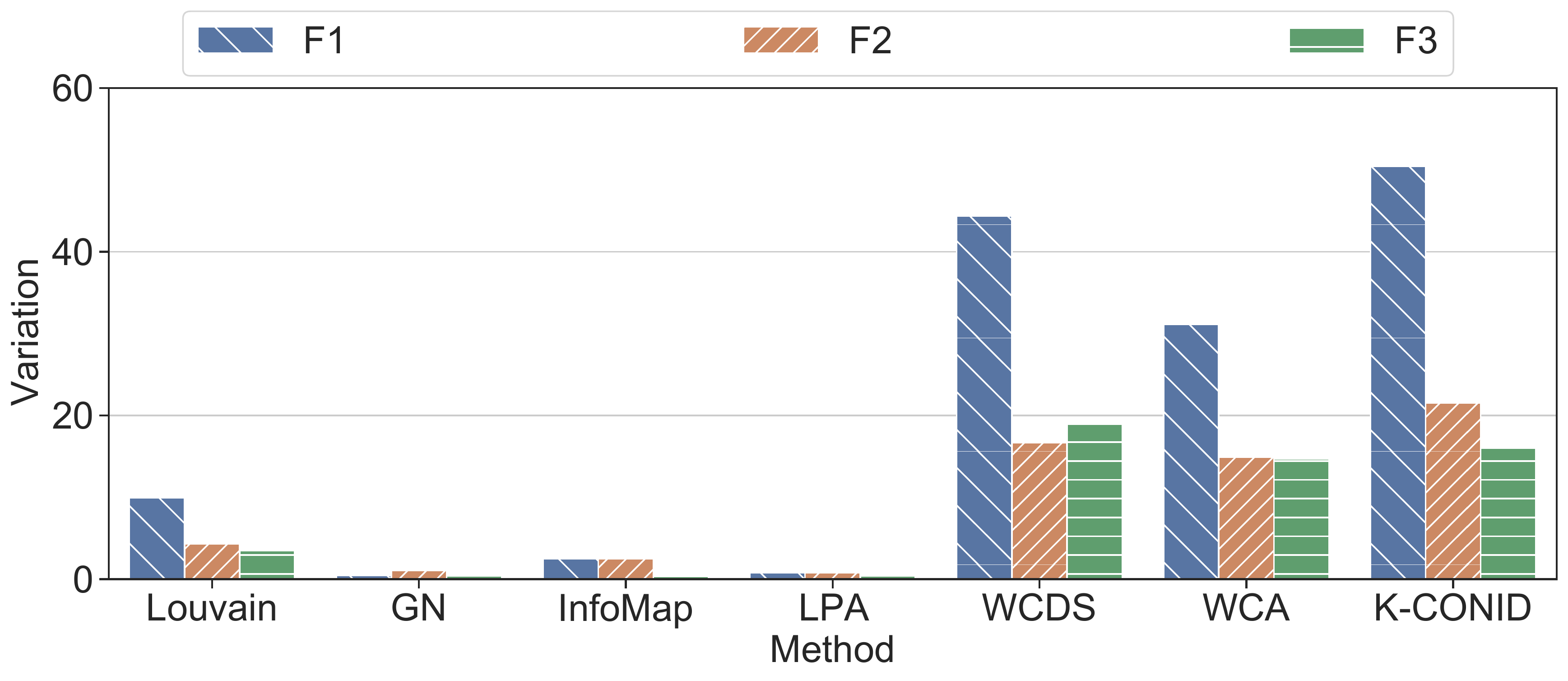}}
\subfigure[Modularity]{
\label{Fig.phase1.2}
\includegraphics[width=0.3\textwidth]{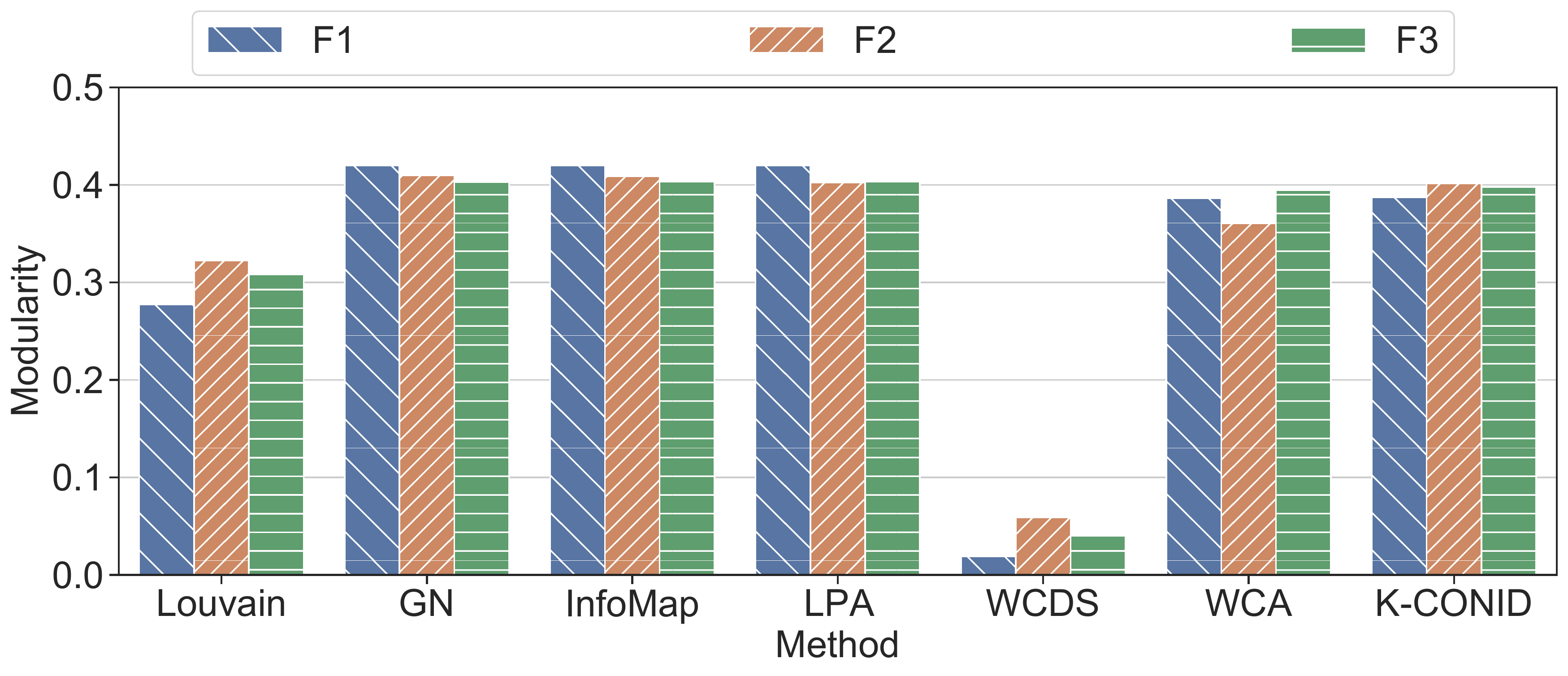}}
\subfigure[Average BER]{
\label{Fig.phase1.3}
\includegraphics[width=0.3\textwidth]{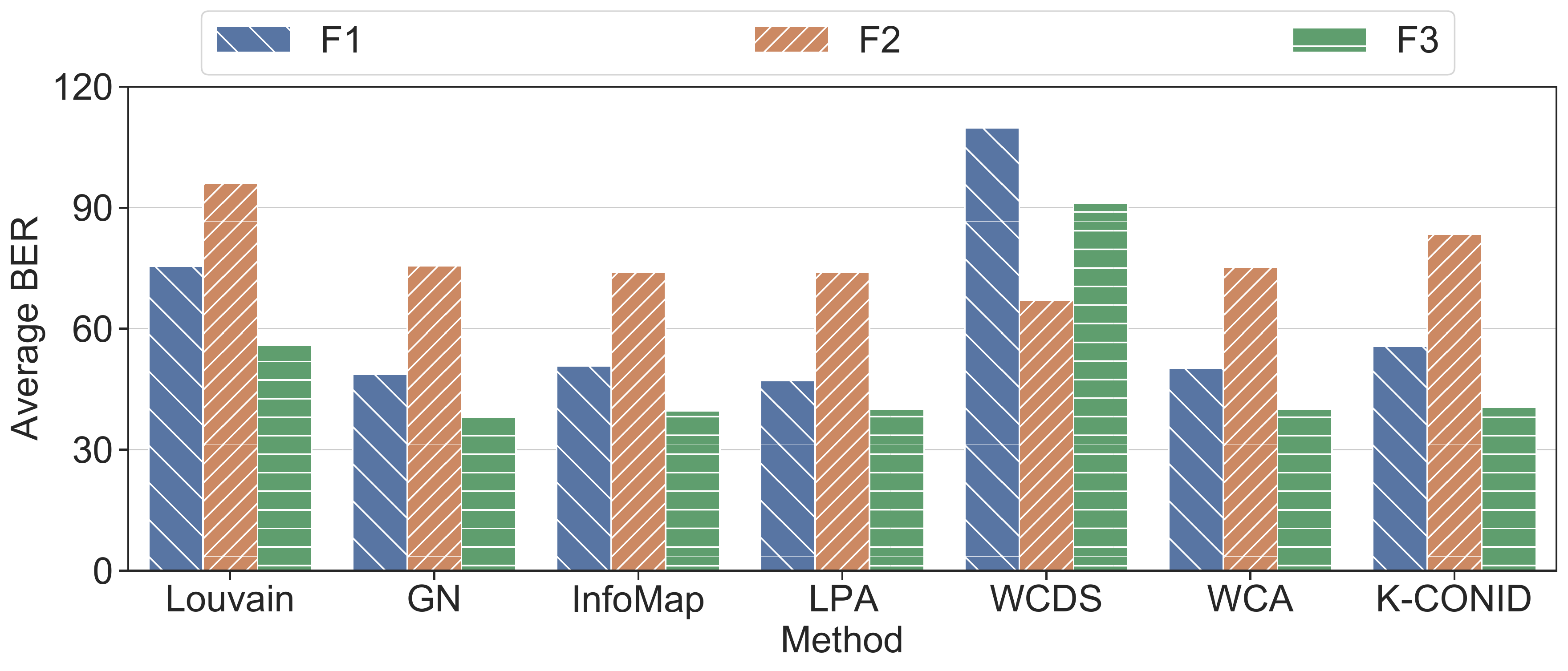}}
\vspace{-1.0em}
\caption{Evaluation Results for Situation \ref{sit:1}}
\label{fig:method1}
\vspace{-1.0em}
\end{figure*}

\begin{table}[htbp]
  \centering
  \caption{Dataset for Situation \ref{sit:1}}
  \vspace{-1.0em}
    \begin{tabular}{cccc}
    \toprule
    Name  & Node  & Total Edge & Timestamp \\
    \midrule
    $F1$    & 62    & 267464 & 391 \\
    $F2$    & 27    & 182748 & 1773 \\
    $F3$    & 29    & 392880 & 2034 \\
    \bottomrule
    \end{tabular}%
  \label{tab:freq}
  \vspace{-1.0em}
\end{table}%

\subsection{Experiment for Situation \ref{sit:1}}
\label{sec:exp:1}


\subsubsection{Evaluation Methods}
\label{sec:exp:1:eva}


The first evaluation method is based on real-life application. From the perspective of daily maintenance and data analysis, a stable partition of network is needed. Thus, we define the \textbf{\emph{variation}} between the partition in two adjacent timestamps. The distance is defined with the edit distance between nodes.
The second evaluation method is based on Modularity. Modularity is a standard to evaluate the result of community detection methods. From the perspective of community detection, modularity evaluates the density of the community. We calculate the average \textbf{\emph{modularity}} in the graphs generated in Step \ref{sec:method:1:community} to compare the performance of the three algorithms.
The third evaluation method is \textbf{\emph{Average BER}}. We remove the inter-community connections, and calculate the shortest distance from one node to another. Average BER is defined as the average distance within all connected components in all graphs in the time window without inter-community edges. The edge weight is defined with the original BER instead of the normalized BER in Section \ref{sec:method:1:graph}.


\subsubsection{Result Analysis}
\label{sec:exp:1:ana}

We compare methods for Situation \ref{sit:1} with conventional clustering methods for ad-hoc networks, including WCDS \cite{WCDS}, WCA \cite{WCA}, and K-CONID \cite{kconid}. 
We examine all seven methods for datasets in Table \ref{sec:exp:dataset}. Figure \ref{Fig.phase1.1} shows the great advantage in our method compared with conventional methods. In the experiment, the time window is fixed to 2. Later research shows that the smaller the time window is, the more stable the partition is. Therefore, the variation in our methods is the worst case among all time windows, but they still outperform the conventional clustering algorithms.
We also test the average modularity on multiple timestamps in Figure \ref{Fig.phase1.2}. Most of our methods outperforms conventional clustering methods on modularity, since the latter ones concentrate on the topology and the degree of nodes, while the community detection methods take advantage of the connectivity and relationships between nodes. However, the performance of Louvain is not as good as other community detection methods, since it is a greedy approach. In addition, we notice that Louvain may generate fake community by splitting one large community into several pieces.
Results for Average BER is shown in Figure \ref{Fig.phase1.3}. The advantage of community detection of our methods is not obvious on this factor. GN reaches the best result in Average BER, but its difference between InfoMap and LPA is subtle. Although the GN algorithm performs better on the data results, we noticed an extremely unbalanced result with a community with only three nodes. Compared to GN, the other three community detection methods can reach more balanced results.

\begin{figure}[htbp]
\vspace{-1.0em}
\centering 
\includegraphics[width=0.3\textwidth]{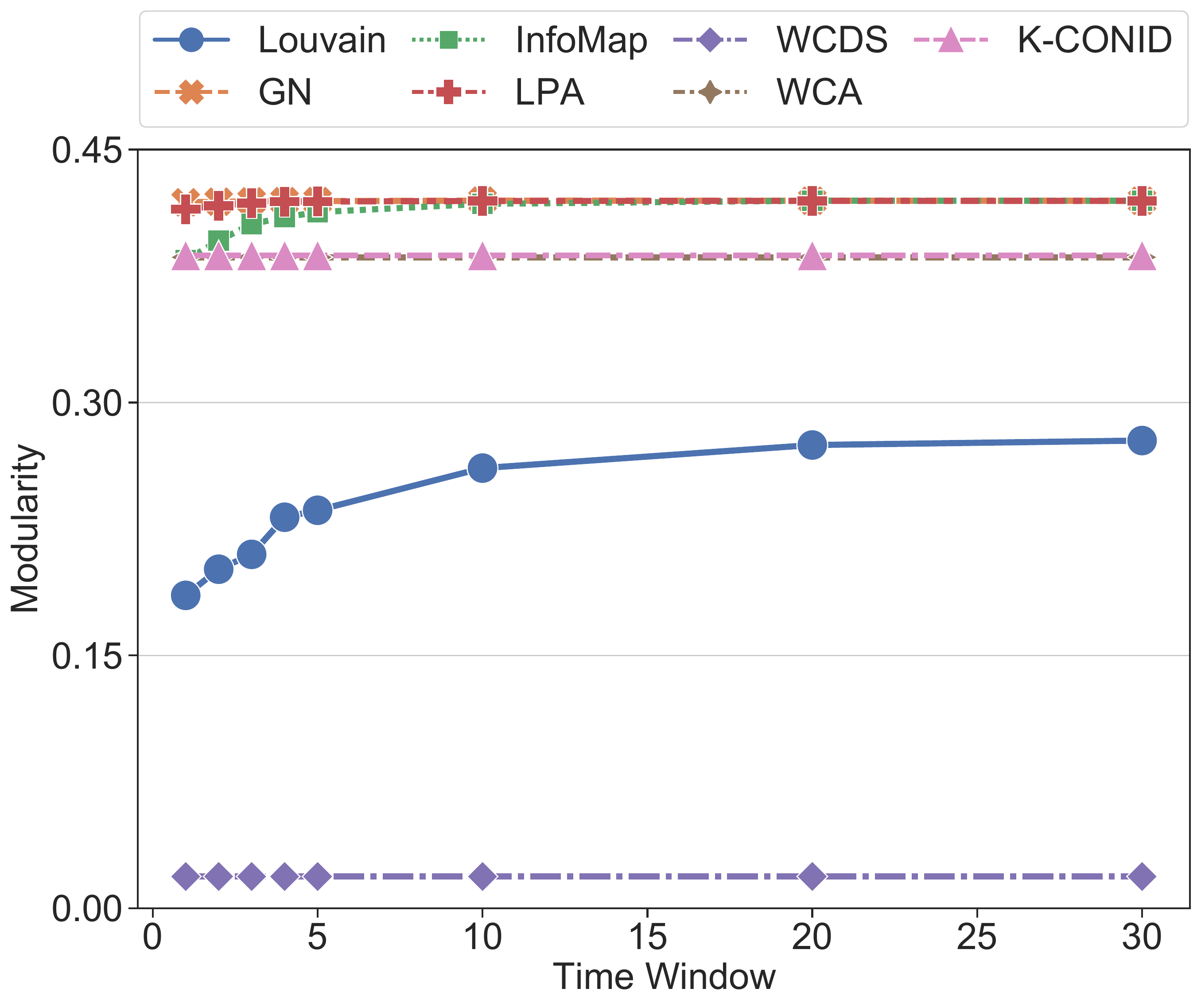}
\caption{The Effect of Time Window on Modularity} 
\label{Fig.window} 
\vspace{-1.0em}
\end{figure}

Finally, we look into the sensitivity of modularity to the size of time window. In Figure \ref{Fig.window}, we show the effect of size of time window on the modularity result. It is obvious that all community detection algorithms converge before the size of window reaches 20, which indicates that in real world application, the size of time window can be controlled without affecting the performance of community detection.

\subsection{Experiment for Situation \ref{sit:2}}
\label{sec:exp:2}

Experiment for Situation \ref{sit:2} is based on a small dataset with 62 timestamps and 3 channels. The evaluation is based on the average modularity and the quality of the partition result. 

\begin{table}[htbp]
  \centering
  \caption{Average BER for Situation \ref{sit:2}}
  \vspace{-1.0em}
    \begin{tabular}{c|cccc}
    \toprule
   Methods & LART  & MNLPA & MultiTensor & InfoMap \\
    \midrule
    Average BER & 74.29 & 45.14  & 68.37 & 59.89 \\
    \bottomrule
    \end{tabular}%
  \label{tab:ber}%
  \vspace{-1.0em}
\end{table}%

Table \ref{tab:ber} shows Average BER result for multiplex community detection. Among the four algorithms, the difference in Average BER is evident. However, in the result of MultiTensor, there exist multiple isolated components.Although MNLPA did best in Average BER, the size of communities are not balanced. It gave only two communities and the difference in size of community is huge. In the result of LART, there are two isolated nodes who is identified as a community, which makes the partition meaningless and should not be accepted. 
To conclude, InfoMap is the best choice for partitioning in ad-hoc networks with multiple channels in our dataset.

\begin{figure}[h]
\vspace{-1.0em}
\centering 
\subfigure[Multiplex Community Detection Result]{
\includegraphics[width=0.23\textwidth]{figures/test1.pdf} 
\label{Fig.method2.1}}
\subfigure[Final Allocation by InfoMap]{
\includegraphics[width=0.23\textwidth]{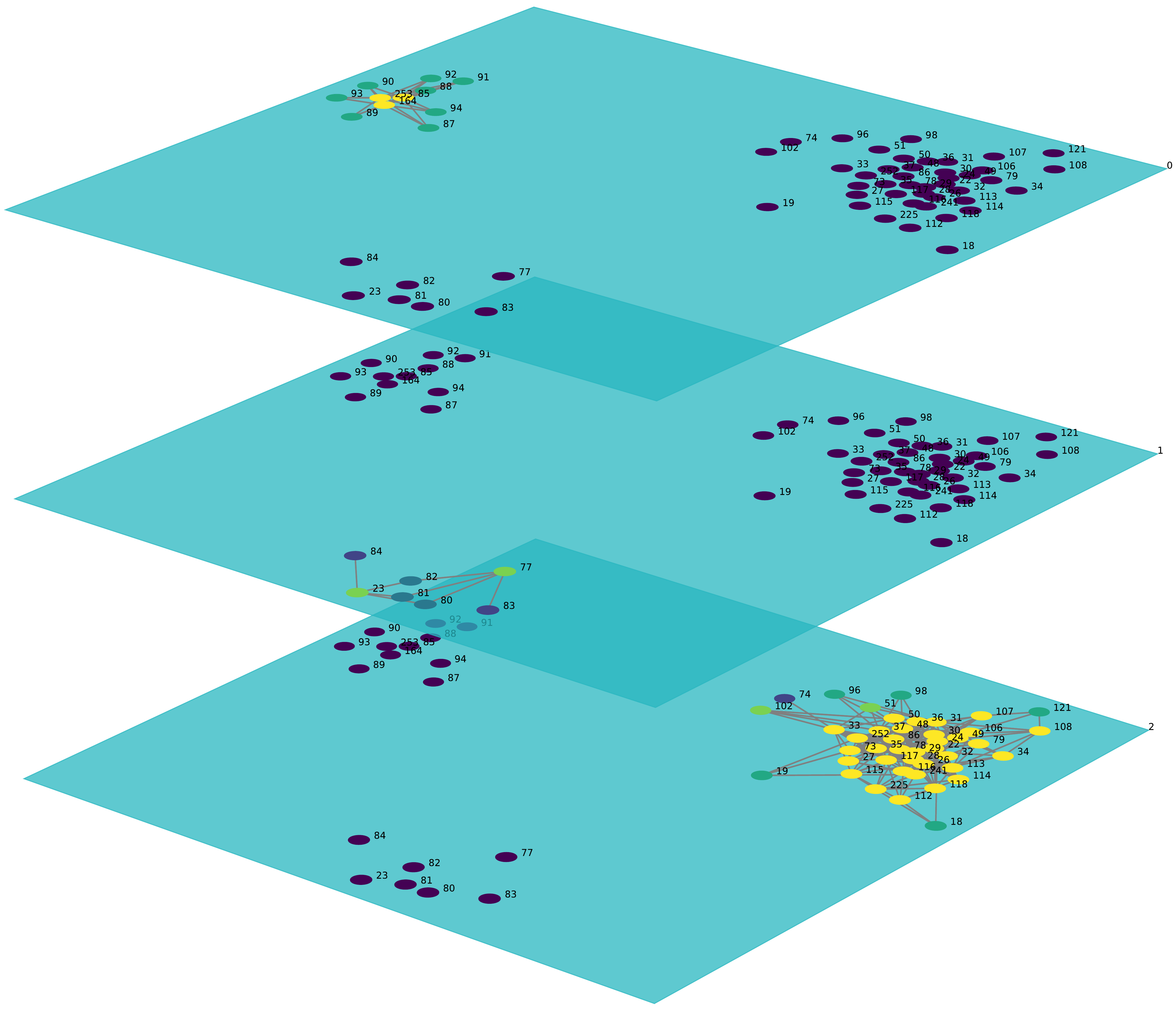} 
\label{Fig.method2.2}}
\label{Fig.Method2}
\vspace{-1.0em}
\caption{Allocation Result for Example \ref{example:1}}
\vspace{-1.0em}
\end{figure}

\begin{example}
\label{example:1}
%
Result Analysis from a real-world dataset  in industrial IoT domain:
The result of multilayer community detection is shown in Figure \ref{Fig.method2.1}. Each color represents a community. It is obvious that the community on three different layers is identical even if the network structure is different. The result of the partition is shown in Figure \ref{Fig.method2.2}, the nodes with colors are the result of the final partition. The two yellow isolated nodes are on layer 3, connected to the red community. In this example, the greedy algorithm gave the optimal result.

\end{example}
\section{Conclusion}
\label{sec:conclusion}

In this paper, we introduce community detection to solve the problem of partition in complex ad-hoc networks. We propose two methods. One is for a series of networks, which calculates the result of community detection to generate similarity and further partition. The other one is for networks with multiple channels, which takes advantage of multilayer community detection algorithms to make a partition for sensors instead of nodes. 
We test our proposals on real-life datasets and choose the most suitable algorithm. One of the advantages of our methods is that each component can be flexibly replaced according to specific application scenarios. This makes our framework more robust.
However, future work is required since the dataset we adopted was relatively small. The robustness should be tested with larger and more complex datasets. In addition, future work should be done to take full advantage of the features of the time series. Further comparisons should also be conducted to find the optimal algorithm apart from the ones listed in this paper. 

\section*{AcknowledgementS}
The work of Y. Wu is supported in part by the National Key R\&D Program of China under Grant 2018YFB1801102 and National Science Foundation(NSFC) under Grant 62071289.

\bibliographystyle{ACM-Reference-Format}
\bibliography{reference}

\end{document}